\documentclass[prl,preprint,showpacs,preprintnumbers,amsmath,amssymb]{revtex4}
\usepackage{color}
\usepackage[dvips]{graphicx}
\DeclareGraphicsExtensions{.eps}
\begin{document}
\preprint{PRL/version1sbw/softoo}
\newcommand{\lsmo}{La$_{0.5}$Sr$_{1.5}$MnO$_4$}
\title{Direct observation of orbital ordering in La$_{0.5}$Sr$_{1.5}$MnO$_4$ using soft x-ray diffraction\\}
\author{S. B. Wilkins} \altaffiliation[Present address: ]{European
  Commission, Joint Research Center, Institute for Transuranium Elements (ITU),
Hermann von Helmholtz-Platz 1, 76344 Eggenstein-Leopoldshafen,
  Germany and European Synchrotron Radiation
  Facility, Bo\^\i te Postal 220, F-38043 Grenoble Cedex, France}\email{wilkins@esrf.fr}
\author{P. D. Spencer}
\author{P. D. Hatton}
\affiliation{Department of Physics, University of Durham,
Rochester Building, South Road, Durham, DH1 3LE, UK}
\author{S. P. Collins}\altaffiliation[Present Address: ]{Diamond Light
  Source Ltd., Rutherford Appleton Laboratory, Chilton, Didcot, Oxon,
  OX11 0QX, UK}
\author{M. D. Roper}
\affiliation{Daresbury Laboratory, Warrington, Cheshire, WA4
4AD, UK}
\author{D. Prabhakaran}
\author{A. T. Boothroyd}
\affiliation{Department of Physics, University of Oxford,
Clarendon Laboratory, Parks Road, Oxford, OX1 3PU, UK}
\date{\today}
\begin{abstract}
We report the first direct resonant soft x-ray scattering observations of
orbital ordering. We have studied the low
temperature phase of La$_{0.5}$Sr$_{1.5}$MnO$_4$, a compound that
displays charge and orbital 
ordering. Previous claims of orbital ordering in such materials have
relied on observations at the Manganese $K$ edge. These claims have
been questioned in several theoretical studies. Instead we have
employed resonant soft x-ray scattering at the manganese $L_{III}$ and $L_{II}$
edges which probes the orbital
ordering directly. Energy scans at constant wavevector are compared to
theoretical predictions and suggest that at all temperatures there are two
separate contributions to the scattering, direct orbital ordering and strong
cooperative Jahn - Teller distortions of the Mn$^{3+}$ ions. 
\end{abstract}

\maketitle

Nearly 50 years ago Goodenough \cite{Goodenough} proposed that
perovskite manganites such as
La$_{1-x}$Ca$_{x}$MnO$_3$ would be orbitally ordered. He proposed that the low
temperature insulating phase contains two sublattices resulting from
charge ordering between Mn$^{3+}$ and Mn$^{4+}$ ions, accompanied by
orbital ordering, whereby the $3d_{3z^2-r^2}$ Mn$^{3+}$ 
orbitals [associated with the long Mn$^{3+}$-O bonds in the Jahn-Teller
  distorted Mn$^{3+}$O$_6$ octahedra] in the form of zigzag chains. This
ordering would entail concomitant displacements of the Mn$^{4+}$O$_6$
octahedra. Since then various crystallographic studies have established
the correlation between, structural, magnetic and transport properties in this and
other related systems. They have also detailed the charge and spin ordering
observed at low temperatures as well as the correlated Jahn-Teller 
distortions. Until recently however, very little was known about the
role of orbital ordering.  This situation has changed dramatically in
the last few years following the claim of direct detection of orbital
ordering by resonant x-ray scattering techniques. Murakami \emph{et al.}\cite{Murakami} 
employed anomalous scattering to observe orbital ordering in the
single layered manganite La$_{0.5}$Sr$_{1.5}$MnO$_{4}$. By tuning to the Mn$^{3+}$ $K$ edge
they observed a sharp reflection at $(\frac{3}{4}, \frac{3}{4},0)$ - a
supercell reflection which
would normally be of zero intensity if the anomalous part of the
scattering factor were isotropic. Moreover, the observation of such
reflections is ruled out by the crystal symmetry of the high
temperature structure, and can only proceed if the symmetry is reduced
by some mechanism, such as the Jahn - Teller effect. Such resonant
techniques have since been 
used to claim orbital order in a number of manganites such as LaMnO$_3$
\cite{Murakami2} and other materials, based on the assumption that
anisotropy in the Mn $4p$ band, probed by $1s - 4p$ excitations at the
Mn $K$-edge, arises from interaction with the anisotropic 3d
states. The claim that such techniques offer a direct probe of orbital
ordering has attracted considerable criticism and 
debate \cite{Ishihara,Elfimov,Benfatto,Mahadevan}. Theoretical studies have
shown that a more likely 
explanation is that the non-isotropic resonance effect is caused by
the oxygen displacements related to the cooperative Jahn-Teller
effect. Recently a theoretical paper suggested the use of resonant x-ray
scattering at the $L_{III}$- and $L_{II}$-edges of Mn$^{3+}$
\cite{Castelton}. Studies at these energies would directly
probe the $3d$ band and it was shown that the shape of the energy
resonances would provide information on whether such ordering was
caused by direct orbital ordering or cooperative Jahn-Teller
displacements.

In a previous paper \cite{Wilkins} we demonstrated how it was possible to use
resonant enhancement at the $L$-edge of manganese to observe
antiferromagnetic spin ordering at low temperatures. In this letter we
report the first unambiguous observation of orbital ordering. We have
conducted the measurements of orbital ordering at the $L_{III}$ and
$L_{II}$ edge of  
Mn$^{3+}$ in La$_{0.5}$Sr$_{1.5}$MnO$_4$. Diffraction experiments have
been undertaken on an
orientated single crystal using a diffractometer placed in a vacuum
vessel to minimize air and window absorption of the soft x-rays
($\approx 640$~eV). Strong resonant enhancement of an orbital order
reflection at  $(\frac{1}{4}, \frac{1}{4},0)$ was observed at the Mn$^{3+}$ edge.  Energy scans through the $L_{III}$
and $L_{II}$ edges were undertaken at constant wavevector at temperatures
up to, and through, the charge and orbital ordering temperature.  We
will demonstrate by using the specific theoretical predictions of
Castleton and Altarelli \cite{Castelton} that the orbital ordering is caused by
Jahn-Teller displacements.

\begin{figure}\begin{center}
\includegraphics[width=0.75\columnwidth]{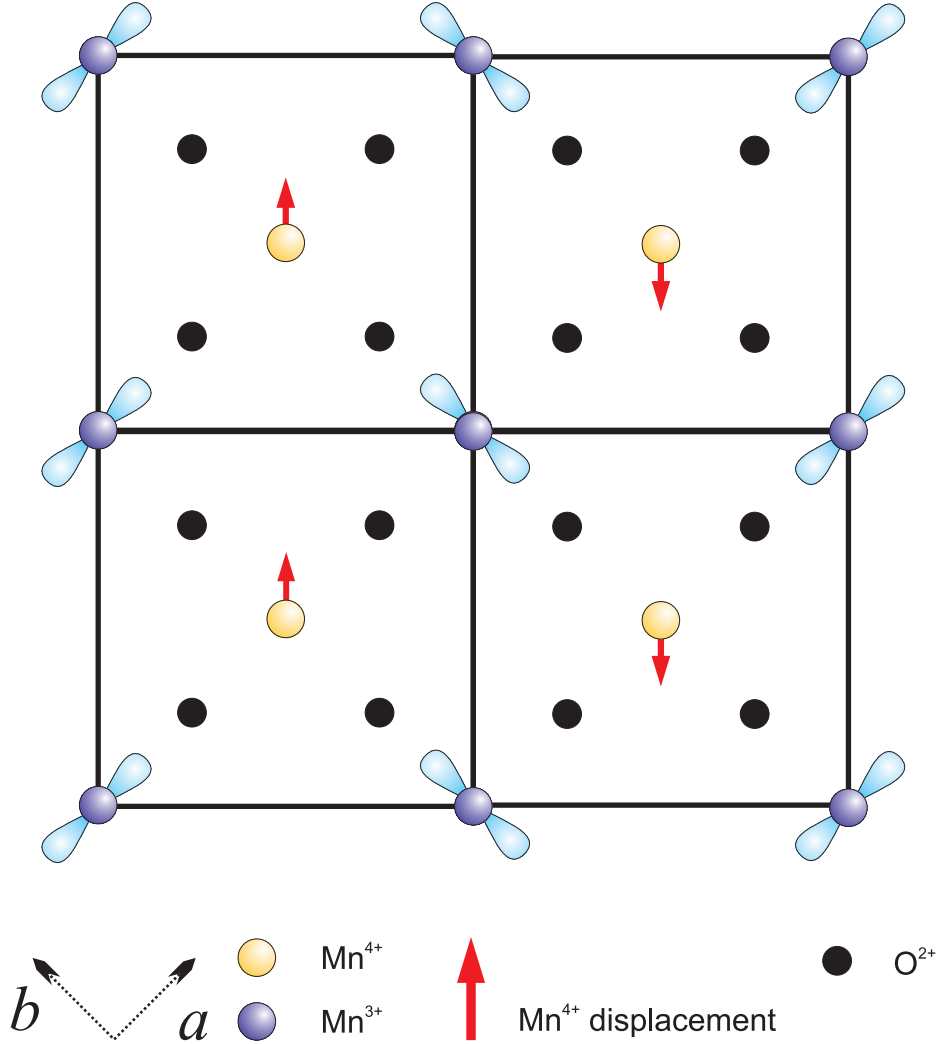}
\caption{Schematic representation of the proposed orbital order
  pattern in th $a-b$ plane.}\label{fig:oo-diagram}
\end{center}\end{figure}

La$_{0.5}$Sr$_{1.5}$MnO$_4$ is a typical transition metal oxide displaying strongly
correlated electron states. Upon cooling the material displays a phase
transition at $T_{CO} = 240$~K into a charge ordered state in which there
is a unit difference in the valence states between subsequent
manganese atoms (see Figure~\ref{fig:oo-diagram}). In addition to the structural Bragg reflections of
the high temperature $I4/mmm$ phase ($a = b = 3.86$~\AA, $c = 12.40$~\AA) other
reflections at a wavevector of  $(\frac{1}{2}, \frac{1}{2},0)$ have been observed which are
a result of this charge ordered state that doubles the unit cell
\cite{Sternlieb}. In addition peaks with a wavevector of $(\frac{1}{4}, \frac{1}{4},0)$ have been
observed in electron diffraction measurements \cite{Moritomo,Bao}. However, subsequent x-ray and neutron
studies have confirmed the presence of the quarter-wave modulation
\cite{Murakami,Larochelle}. This additional wavevector implies that the $a$ and $b$ axes of
this high temperature tetragonal phase are quadrupled. However, a
smaller $A$-centered orthorhombic unit cell, rotated by $45^\circ$
with $a_o \approx b_o \approx 2\sqrt{2}a_t$ and
$c_o = c_t$ can be used to index all the reflections. The nature of this
additional modulation is still unclear. $K$ edge resonant x-ray
scattering was used to claim that this quarter-wave modulation was due
to orbital ordering \cite{Murakami}. Another study using non-resonant techniques
suggested that this could be caused by cooperative Jahn-Teller
distortions causing a shear-type distortion, similar to that proposed
in La$_{0.5}$Ca$_{0.5}$MnO$_3$ and LaSr$_2$Mn$_2$O$_7$ \cite{Larochelle}. Irrespective of the mechanism,
it is clear that orbital order develops at the same temperature as the
charge ordering ($T_{OO} = T_{CO}= 240$~K).

Further cooling results in long-range spin ordering. As observed by
neutron scattering \cite{Sternlieb} a complex antiferromagnetic ordering occurs at
$T_N = 110$~K involving both manganese sublattices. Magnetic
susceptibility measurements give a much higher 
antiferromagnetic transition temperature, concurrent with the charge
and orbital ordering \cite{Damay}. It therefore seems likely that in-plane
antiferromagnetic order develops at 240 K ($T_N(ab) = T_{CO} = T_{OO}$)
consistent with the rod-like neutron scattering reported between T$_N$
and T$_{CO}$ \cite{Sternlieb}. This in-plane antiferromagnetic order then develops fully
three dimensionally at $T_N(c) = 160$~K. At the Mn$^{3+}$ sites ($3d^4$) the
Hunds rule coupling is strong and the crystal field has a large cubic
($O_h$) component. This causes a twofold degenerate configuration $t^3_{2g\uparrow}e^1_{g\uparrow}$. This
degeneracy can be lifted by a Jahn-Teller distortion of the MnO$_6$
octahedron reducing the symmetry to $D_{4h}$ and separating the two
components of the $e_{g\uparrow}$ level into $3d_{3z^2-r^2\uparrow}$ and $3d_{x^2-y^2\uparrow}$. Goodenough \cite{Goodenough} showed that the
spin ordering is dependent upon the ordering of this orbital degree of
freedom and hence a distinctive ``herring-bone'' pattern of the
orbitals is required to explain the observed antiferromagnetic spin
structure. This orbital order pattern has a wavevector of $(\frac{1}{4}, \frac{1}{4},0)$. 

There are thus at least two possible mechanisms to account for the
observed orbital ordering. It could arise from 
spin ordering that forms at $T_N$. Alternatively the orbital ordering
could be a consequence of the cooperative Jahn-Teller that forms at
$T_{CO}$. In some manganites, such as LaMnO$_3$ \cite{Rodriguez} and
La$_{0.5}$Ca$_{0.5}$MnO$_3$ \cite{Radaelli},
the extent of the Jahn-Teller distortions has been observed using
crystallographic refinements of neutron and x-ray diffraction. The
distortions vary between 7 and 12\% suggesting that the Jahn-Teller
mechanism may be important. However in La$_{0.5}$Sr$_{1.5}$MnO$_4$ the MnO$_6$
octahedra appear to be almost undistorted with a fractional change in
the Mn-O bond length of only $\approx1$\% \cite{Sternlieb}. This led
to the suggestion that
in La$_{0.5}$Sr$_{1.5}$MnO$_4$ the principle mechanism was the Goodenough spin
mechanism. 

In order to understand the interaction and interdependence
of the charge, spin, orbital and Jahn-Teller distortion degrees of
freedom it is vital to develop techniques that can distinguish between
orbital and cooperative Jahn-Teller ordering. The $K$ edge experiment of
Murakami \emph{et al.}, is only
an indirect indication of $3d$ orbital ordering as they probe
the $4p$ shell. Any sensitivity of the $4p$ band to the splittings in the
$3d$ electron band was believed to arise from a combination of Coulomb
interaction with the ordered $3d$ electrons and Jahn-Teller distortions
\cite{Ishihara}. However subsequent theoretical studies have shown
that $K$ edge
experiments are about 100 times more sensitive to Jahn-Teller
distortions than to orbital ordering and cannot be said to probe 3d
ordering directly.

The obvious method is resonant scattering at the $L$
edges, which directly probe the $3d$ shell itself. Unfortunately the
$L_{III}$ and $L_{II}$ edges of manganese are in the soft x-ray region with
energies of only $\sim650$~eV. Diffraction experiments using such
wavelengths are difficult and only very recently we reported
the first diffraction experiments at the $L$ edges of manganese from a
bulk single crystal of La$_{1+2x}$Sr$_{2-2x}$Mn$_2$O$_7$ ($x = 0.475$)
\cite{Wilkins}. 

\begin{figure}\begin{center}
\includegraphics[width=\columnwidth]{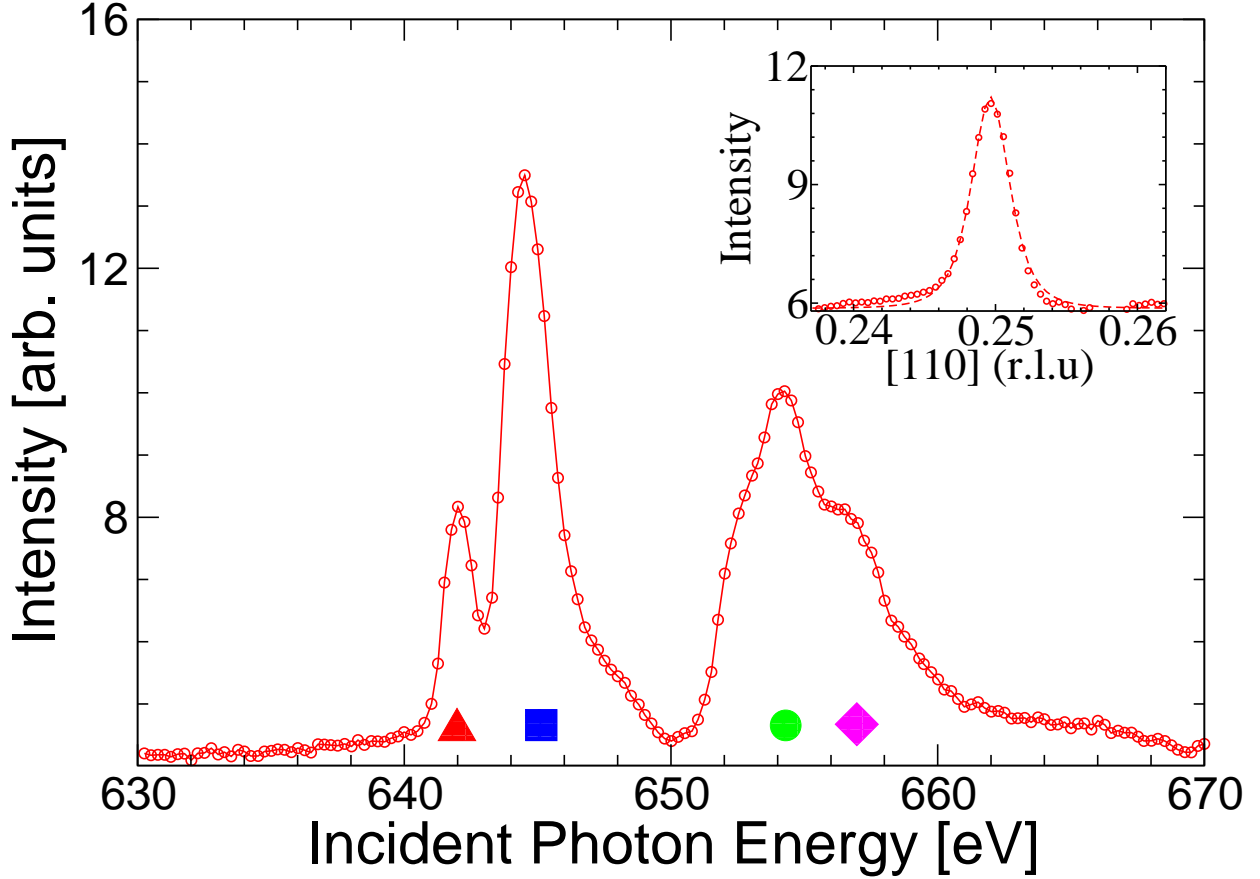}
\caption{Energy scan at fixed wavevector through the $(\frac{1}{4},
  \frac{1}{4}, 0)$ orbital order reflection at 140~K. The symbols
  ($\blacktriangle,
  \blacksquare, \bullet, \blacklozenge$) show the energies of the four
  major features in the scan as in Ref.\cite{Castelton}. The inset shows a scan
  taken at $E = 641$~eV through the orbital order reflection in the
  $[110]$ (longitudinal) direction. 
By fitting the peak to a Lorentzian squared lineshape, the inverse correlation length (defined as 
  $\zeta^{-1}=\frac{2\pi}{c}\kappa$, where 
$\kappa$ is the half width at half maximum in reciprocal lattice units
(r.l.u.) and $c$ is the direct space lattice parameter) is found to be $2.6
  \times 10^{-3}$~\AA$^{-1}$. 
}\label{fig:exp-data-resonance}
\end{center}\end{figure}

The present experiments were conducted on station 5U1 at the
Synchrotron Radiation
Source (SRS) at Daresbury Laboratory. A single crystal of
La$_{0.5}$Sr$_{1.5}$MnO$_4$ was grown by the floating zone method with dimensions
of $4 \times 4 \times 1$~mm. The sample was cut and polished with $[110]$ surface normal, and  aligned
in the two-circle diffractometer \cite{diffo} with the $[110]$ and $[001]$
axes defining the diffraction plane. The beamline, situated at the end of a
variable gap undulator, produces a tuneable monochromatic beam of
photons with an energy resolution of $\Delta E = 1$~eV at 640~eV. The incident photon energy was tuned to the manganese $L_{III}$ edge
(639~eV) and a search was undertaken for a superlattice reflection at
 $(\frac{1}{4}, \frac{1}{4},0)$ at 85~K. At 639~eV the reflection would be expected to occur
at a Bragg angle of $\theta = 62.9^\circ$. A clear peak was observed at this
position, a scan of which in the longitudinal direction, $Q_z$, is shown
in the inset of Figure~\ref{fig:exp-data-resonance}. The peak was well defined with a Lorentzian squared line
shape which was significantly broader than the instrumental resolution.

An energy scan of this orbital ordering reflection was then undertaken
at constant wavevector. The results of which are shown in Figure~\ref{fig:exp-data-resonance}. The orbital ordering reflection was only observed
close to the L$_{III}$ and $L_{II}$ edges. No scattering above background was
observed at the $L_I$ edge or at any other energy away from the $L_{III}$ and
$L_{II}$ edges. No resonant scattering was observed at the oxygen $K$ edge or
the lanthanum $M$ edges. This demonstrates the extreme atomic
selectivity of such measurements. We are therefore confident that
these effects are dominated by ordering of the Mn$^{3+}$
sublattice. The interpretation of the energy scan of Figure~\ref{fig:exp-data-resonance} relies
on a series of theoretical calculations and predictions by Castleton
and Alterelli \cite{Castelton}. In their paper they made specific predictions based
on atomic multiplet calculations as to the energy dependence of the $L$
edge orbital ordering reflection in La$_{0.5}$Sr$_{1.5}$MnO$_4$. They even
calculated the different types of resonance that would be observed if
the orbital ordering originated from a direct Goodenough model, or as
a consequence of cooperative Jahn-Teller distortions. 

\begin{figure}\begin{center}
\includegraphics[width=\columnwidth]{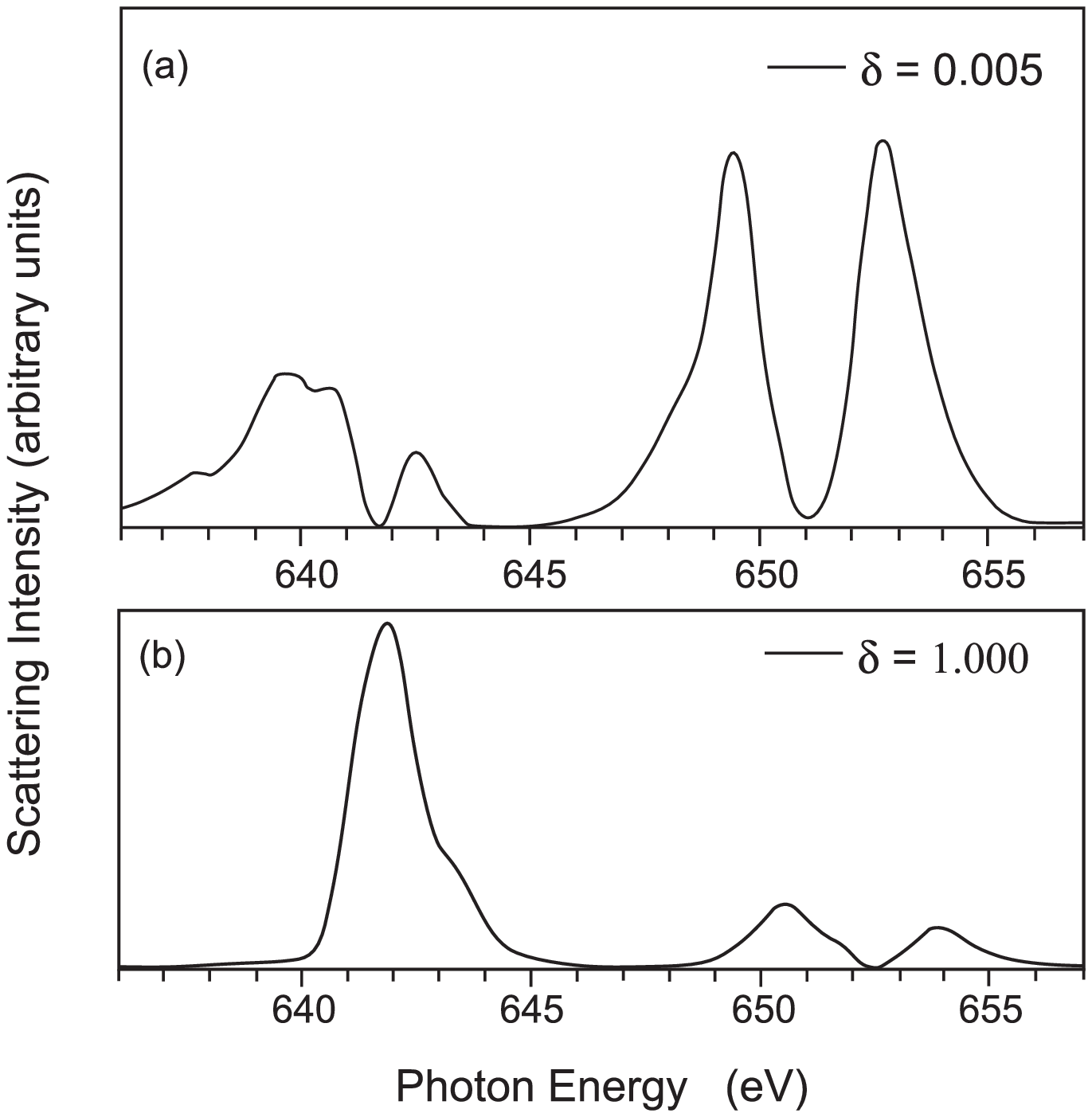}
\caption{Energy spectra prediction by Castleton and
  Altarelli\cite{Castelton} for the $(\frac{1}{4},\frac{1}{4},0)$
  reflection with $\delta = 0.005$ (no Jahn - Teller) and $\delta = 1$
  (Strong Jahn-Teller).}\label{fig:castelton_prediction}
\end{center}\end{figure}

The energy scan of Figure~\ref{fig:exp-data-resonance} displays four clear features of differing
widths and intensities. The calculations of Castleton and Altarelli
predict four major features, which were interpreted within a simple
one-electron picture of Mn$^{3+}$. The energy dependence carries specific
information regarding the type and origin of the orbital ordering. In
the $K$ edge experiments of Murakami \cite{Murakami} it was impossible to
differentiate between the $3d_{3z^2-r^2} / 3d_{3x^2-r^2}$ ordering and
the alternative $3d_{3z^2-r^2} / 3d_{x^2-y^2}$ ordering. However Castleton and Altarelli \cite{Castelton} include in their paper
calculations for both types of orderings that are very different. A
brief comparison of our data and their calculations would suggest that
the orbital ordering in La$_{0.5}$Sr$_{1.5}$MnO$_4$ consists of the latter
type. However at this stage some degree of caution is required, as the
calculations are not derived from experimental spectra. Further work
is therefore required involving detailed fitting to the data before a
definitive answer can be given. Our work however does indicate, even
at this early stage, the additional information available from direct
measurements at the $L_{III}$ and $L_{II}$ edges.

The energy scan of Figure~\ref{fig:exp-data-resonance} does however give definitive answers as to
the mechanism of the orbital ordering. As discussed earlier the
observed scattering could be due to direct Goodenough orbital ordering alone, or could be a
consequence of strong cooperative Jahn-Teller distortions on the Mn$^{3+}$
ions which causes the orbital ordering. Again, Castleton and Altarelli \cite{Castelton} produced detailed predictions
as to the energy spectra that would be observed for each limiting case,
and even for a mixture of both via a scaling parameter, $\delta$.
$\delta = 0$ effectively turns off the lattice distortion and simulates
the effect of pure orbital ordering and no Jahn-Teller
distortions. The predicted energy dependence for $\delta = 0.005$ is
given in Figure~\ref{fig:castelton_prediction}b. For such a case, and
even for the case of the 
Jahn-Teller distortion being relatively small ($0 < \delta < 0.25$) the
predominant intensities should be contained in the higher energy
features associated with the $L_{II}$ edge. For a mixture of orbital
ordering and Jahn-Teller distortions ($0.25 < \delta < 0.75$) the intensity
of the $L_{II}$ peaks is relatively constant but there is a rapid rise in
the intensity of a peak at $\sim642$~eV just above the $L_{III}$ edge. Finally,
for dominant Jahn-Teller distortions ($0.75 < \delta < 1$) the spectral weight
resides in two features close to the $L_{III}$ edge (see
Figure~\ref{fig:castelton_prediction}b). Our observations could not be
clearer. Even in 
La$_{0.5}$Sr$_{1.5}$MnO$_4$, the archetypal material for orbital
ordering in which 
it had been claimed that Jahn-Teller distortions might be relatively
small, Figure~\ref{fig:exp-data-resonance} shows that cooperative Jahn-Teller
distortions are the dominant cause of the observed orbital
ordering. This is simply because of the strong intensity at the $L_{III}$ edge of the
two low-energy features. However, it is also apparent that there is
considerable intensity at the $L_{II}$ edge which can only be caused by direct
Goodenough orbital ordering. We therefore conclude that there are two separate
contributions to the observed orbital ordering, direct Goodenough orbital
ordering and strong cooperative Jahn-Teller distortions of the Mn$^{3+}$
ions. All previous studies have only considered a single contribution to the
orbital order. 

\begin{figure}\begin{center}
\includegraphics[width=\columnwidth]{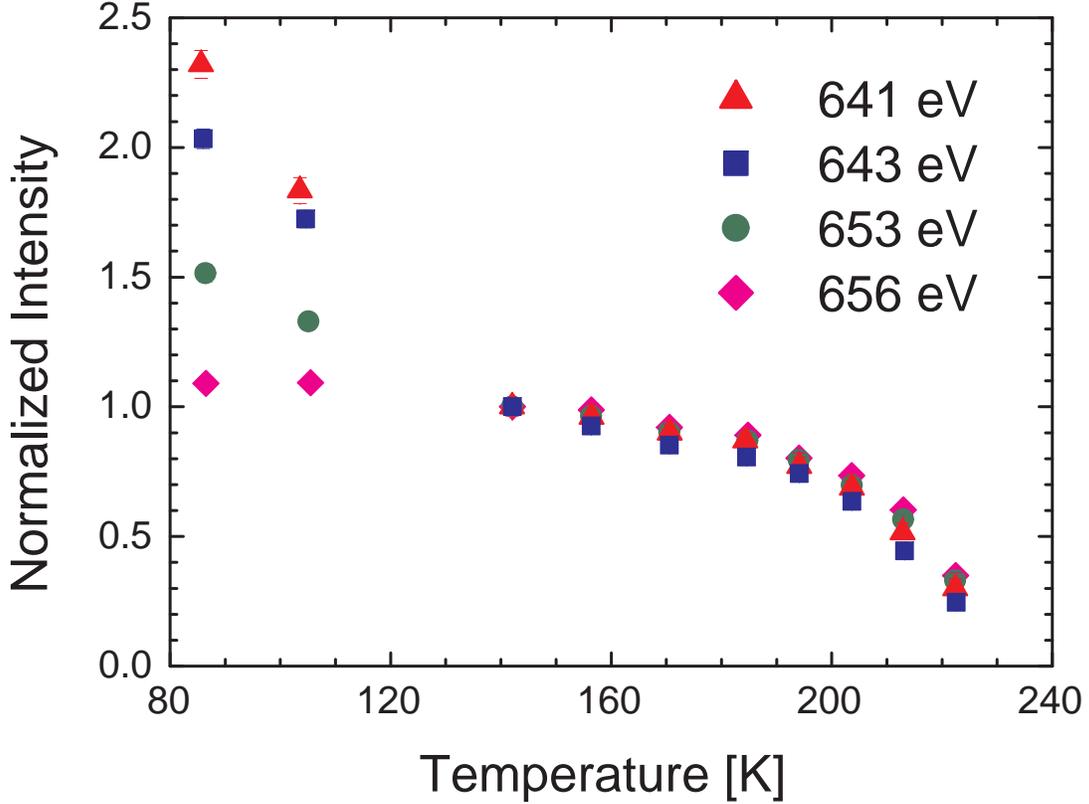}
\caption{Temperature dependence of the integrated intensity (measured in the
  longitudinal direction) of the $(\frac{1}{4}, \frac{1}{4}, 0)$  superlattice
  reflection as a function of temperature of the four features in the energy
  spectrum. The intensities have been normalized at 140~K.}\label{fig:temp-dep}
\end{center}\end{figure}

The temperature dependence of the orbital
ordering below the phase transition was also measured. Energy scans at
constant wavevector were also taken of the orbital ordering peak to
determine if the cause of such orbital ordering had a temperature
dependence. A graph of the integrated
intensity of the four individual energy resonances comprising the
orbital order peak is shown in Figure~\ref{fig:temp-dep}. This shows the 
disappearance of the orbital ordering peak as the sample is warmed up
to and through the orbital ordering phase transition. 
Figure~\ref{fig:temp-dep} shows that the lower energy peaks attributed to
the cooperative Jahn-Teller effect become even more dominant at lower
temperatures. This 
demonstrates that the two causes of the orbital ordering are separate,
because they have a different temperature dependence. Detailed fitting of the
temperature dependence will be undertaken in a future paper but at this stage
we will simply note the differing response to temperature of direct Goodenough
orbital order (the features at 653~eV and 656~eV) and those of Jahn-Teller
distortions (those at 641~eV and 643~eV).

To conclude, our results demonstrate the first use of soft x-ray
diffraction to probe orbital ordering. Resonant scattering at the $L$
edge of manganese provides a direct probe of the orbital order and
provides information on the electronic configuration and the
underlying mechanism of the orbital order. In
La$_{0.5}$Sr$_{1.5}$MnO$_4$ we find 
that both orbital and charge ordering develop at the first-order
structural transition at 240 K. The charge ordering causes a 
configurational ordering of the Mn$^{3+}$ and Mn$^{4+}$ ions into a checker
board pattern. Jahn-Teller distortions of the Mn$^{3+}$ ions causes
considerable distortion of the MnO$_6$ octahedra and the individual Mn-O
bond lengths and angles. This, simultaneously, causes cooperative
displacements of the Mn$^{4+}$ ions and orbital ordering on the Mn$^{3+}$
sublattice. Spin ordering of the Mn$^{3+}$ and Mn$^{4+}$ sublattices
occurs at a 
lower temperature and does not directly cause the observed orbital
ordering. It is likely that the strong cooperative Jahn-Teller
distortions are one of the causes of orbital ordering in \emph{all} the
manganites. Searches for pure Goodenough type orbital ordering caused by
the low temperature spin ordering arrangements should be conducted in
materials that do not display strong Jahn-Teller distortions. Finally
our results show that soft x-ray resonant
diffraction will become a major technique for the study of charge,
spin and orbital ordering in a wide range of materials.

\section{Acknowledgements}

We gratefully acknowledge experimental support and discussions with
T.A.W. Beale. We are
grateful to the director of the Synchrotron Radiation Source for
access to the beamline and to EPSRC and CLRC for financial
support. PDS and SBW would like to thank EPSRC for financial support and PDH
thanks the University of Durham Research Foundation for financial support.


\begin{thebibliography}{99}
\bibitem{Goodenough}J. B. Goodenough \emph{Phys. Rev.} {\bf100}, 564
  (1955)
\bibitem{Murakami}Y. Murakami \emph{et al.,
  Phys. Rev. Lett.} {\bf80}, 1932 (1998).
\bibitem{Murakami2}Y. Murakami \emph{et al.,
  Phys. Rev. Lett.} {\bf81}, 582 (1998).
\bibitem{Ishihara}S. Ishihara and S. Maekawa, \emph{Phys. Rev. Lett.}
  {\bf80}, 3799 (1998).
\bibitem{Elfimov}I.S. Elfimov \emph{et al., Phys. Rev. Lett.} {\bf82},
  4264 (1999).
\bibitem{Benfatto}M. Benfatto, Y. Joly, and C. R. Natoli,
  \emph{Phys. Rev. Lett} {\bf83}, 636 (1999)
\bibitem{Mahadevan}Priya Mahadevan, K. Terakura and D.D. Sarma,
  \emph{Phys. Rev. Lett.} {\bf 87}, 066404 (2001)
\bibitem{Castelton}C. W. M. Castleton and M. Altarelli,
  \emph{Phys. Rev. B} {\bf62}, 1033 (2000).
\bibitem{Wilkins}S. B. Wilkins, P. D. Hatton, \emph{et
  al. Phys. Rev. Lett.} {\bf 90}, 187201 (2003).
\bibitem{Sternlieb}B.J. Sternlieb \emph{et al., Phys. Rev. Lett.}
  {\bf76}, 2169 (1996).
\bibitem{Moritomo}Y. Moritomo \emph{et al., Phys. Rev. B} {\bf51},
  3297 (1995).
\bibitem{Bao}W. Bao \emph{et al., Solid State Commun.} {\bf98}, 55
  (1996).
\bibitem{Larochelle}S. Larochelle \emph{et al., Phys. Rev. Lett.}
  {\bf87}, 095502 (2001).
\bibitem{Damay}F. Damay \emph{et al., J. Magn. Magn. Mater.} {\bf183},
  143 (1998).
\bibitem{Rodriguez}J.Rodriguez-Caravajal \emph{et al., Phys. Rev. B.}
  {\bf57}, R3189 (1998).
\bibitem{Radaelli}P.G. Radaelli \emph{et al., Phys. Rev. Lett.}
  {\bf75}, 4488 (1995).
\bibitem{diffo}M.D. Roper \emph{et al., Nucl. Instrum. Methods
  Phys. Res. Sect. A-Accel. Spectrom.} {\bf}467, 1101 (2001).

\end{thebibliography}
\end{document}